%% file: paper.tex
\documentclass[10pt,conference]{IEEEtran}
\IEEEoverridecommandlockouts

\usepackage{verbatim}%
\usepackage{fancyvrb}%
\usepackage{url}%
\usepackage{comment}%
\usepackage{graphicx}%
\usepackage{multirow}%
\usepackage{float,fancyvrb}%
\usepackage{algorithm}%
\usepackage{algpseudocode}%
\usepackage{amsthm}%
\usepackage{amsmath,amssymb,amsfonts}%
\usepackage{mathrsfs}%
\usepackage{xcolor}%
\usepackage{textcomp}%
\usepackage{mathtools}%
\usepackage{booktabs}%
\usepackage{multicol}%
\usepackage{float}

\newtheorem{theorem}{Theorem}
\newtheorem{lemma}{Lemma}
\begin{document}

\title{Multi-Pass Targeted Dynamic Symbolic Execution
\thanks{This work has been partially funded by the U.S. National Science Foundation through CAREER Award \# 1942235.}}

\author{\IEEEauthorblockN{Tuba Yavuz}
\IEEEauthorblockA{ECE Department\\
University of Florida\\
tuba@ece.ufl.edu
}
}

\maketitle

\begin{abstract}
\input{abstract}
\end{abstract}


\input{introduction}

\input{overview}

\input{approach}
\input{evaluation}

\input{relatedwork}
\input{conclusions}

\input{paper.bbl}

\end{document}

%% file: abstract.tex
Dynamic symbolic execution (DSE) provides a precise means to analyze 
programs and it can be used to generate test cases and to detect a variety of bugs 
including memory vulnerabilities.
However, the path explosion problem may prevent a symbolic executor from covering
program locations or paths of interest.
In this paper, we present a Multi-Pass Targeted Dynamic Symbolic Execution approach 
that starts from a target 
program location and moves backward until it reaches a specified entry point 
to check for reachability, to detect bugs on the feasible paths 
between the entry point and the target, and to collect constraints about the 
memory locations accessed by the code.
Our approach uses a mix of backward and forward reasoning passes. 
It introduces an abstract address space that gets populated during the 
backward pass and uses unification to precisely map the abstract objects to 
the objects in the concrete address space. 
We have implemented our approach in a tool called DESTINA using KLEE, 
a DSE tool. 
We have evaluated DESTINA using SvComp benchmarks from the memory safety and control-flow 
categories.
Results show that DESTINA can detect memory 
vulnerabilities precisely and it can help DSE reach target locations faster when it 
struggles with the path explosion. Our approach achieves on average 4X reduction 
in the number of paths explored and 2X speedup.

%% file: introduction.tex
\section{Introduction}
\label{sec:intro}

Symbolic execution \cite{King76} has become an important technique for program testing and 
bug detection. 
It is a path-precise program analysis approach that uses symbolic inputs and generates a set of paths according to the program semantics.
Each program statement is executed within the context of the concrete address space that can store concrete as well as symbolic values, which enables the propagation of symbolic inputs and expressions to other variables and the generation of a constraint or a path condition over the symbolic inputs. 
Branch conditions involving symbolic expressions lead to forking, which generates a 
clone of the path that differs from the original one based on the branch target taken.
The feasibility of the clone is checked using an SMT solver. 
In symbolic execution, the number of paths generated can quickly grow when there are nested control-flow statements or recursion that are controlled by symbolic constraints leading to the well-known path explosion problem.
Symbolic execution can provide useful results in the form of tests and bug reports 
before the path explosion happens. There has been a plethora of work that deals with the 
path explosion problem in the context of symbolic execution through 
path exploration heuristics and guiding the execution towards a specific code location or a goal \cite{MK11,DA14,YYG15,GKW15,YYG18,TMR18,CS21,ZCW22}, which we call targeted symbolic execution. Existing targeted symbolic execution approaches do not support backward execution with the exception of \cite{DA14,CS21}, which do not support pointer arithmetic 
and aliasing.

In this paper we present a multi-pass targeted symbolic execution approach that 
can start the exploration from a target location and condition\footnote{As long as the condition can be specified as a condition in terms of program variables.}.  
We have implemented our approach in a tool called DESTINA\footnote{We will release DESTINA on github.}, which is built on top of KLEE \cite{klee}.
It creates an abstract address space and uses it during the backward pass 
to record the side effect of the executed statements in terms of abstract expressions. 
It keeps track of the points-to information for arbitrary pointer expressions and discovers the connections between the abstract objects and the concrete objects 
as it soundly establishes the {\tt def-use} dependencies.
It can detect the infeasibility of a path before reaching the entry location by 
resolving abstract objects using objects from the concrete address space.
Backward pass also handles the forking of paths by interpreting the control-flow in 
backward direction and recording the constraints for the path condition.
The backward pass is followed by a forward pass that propagates 
the unification of abstract objects with the concrete ones and resolves 
expressions by rewriting them using concrete values or addresses of the concrete objects.  

DESTINA uses byte precise memory modeling to handle memory accesses in a precise way.
This enables DESTINA to generate an upside down symbolic execution tree, where 
the path condition of each path consist of a set of resolved and unresolved 
constraints. If a subset of the resolved path constraints become unsatisfiable, 
DESTINA declares the path to be infeasible and prunes it.
DESTINA is equipped with heuristics to deal with the path explosion and improve its performance. Specifically, it can collect constraints even from infeasible paths to guide standard dynamic symbolic execution towards the target.

To our knowledge, DESTINA is the first targeted symbolic execution approach that 
can analyze programs with pointer arithmetic and dynamic memory management, and, therefore, 
it can precisely detect memory vulnerabilities including NULL pointer dereferences, out-of-bound memory accesses, double-frees, use-after-frees, and memory leaks.


We answer the following research questions in this paper:
\begin{itemize}
    \item {\bf RQ1:} Can DESTINA detect memory errors precisely? 
    \item {\bf RQ2:} Can DESTINA guide DSE to reach the target locations faster?
\end{itemize}

Our results on SvComp \cite{svcomp} benchmarks from the memory safety and control-flow categories show that DESTINA can detect a variety of memory vulnerabilities precisely and quickly
and it can help  DSE reach the targets on average 2X faster while achieving  a 4X reduction in the number of paths explored. We will release DESTINA on github.


%% file: overview.tex
\section{Overview}
\label{sec:overview}

\begin{figure}[th!]
\begin{footnotesize}
\begin{Verbatim}[numbers=left,xleftmargin=5mm]
     i = 0;
     W = newSymbolic;
     while (i<W)
        i++;
     if (W == 1000) {
        p = NULL;
        *p = 7;      
        r = *p;
        ...
       if (Y) foo(); 
       else bar(); 
     }
\end{Verbatim}
\end{footnotesize}
\caption{A path explosion example with a NULL pointer dereferencing error. {\tt W} and {\tt Y} are symbolic.}
\label{fig:needforguidance}
\end{figure}

In this section, we motivate our targeted dynamic symbolic execution approach. The main challenge in dynamic symbolic execution is 
the path explosion problem, which represents an exponential increase in 
the number of paths generated by a symbolic executor. 
Such an increase may prevent the symbolic executor from achieving its goal,  
which may include achieving a certain level of code coverage or  
finding bugs in a reasonable amount of time.
In this paper, we assume that the analyst has a goal in the form of a target, i.e., a 
program location of interest, and wants to find out about the target's reachability and 
the errors that get manifested on the paths that reach the target. 
Below, we discuss several examples to motivate the problem.

Figure \ref{fig:needforguidance} shows an example to explain the need for a targeted analysis. The while loop at lines 3-4 runs {\tt W} times, which gets assigned a symbolic value at line 2. The {\tt if} condition at line 5 evaluates to true 
only after the loop body at line 4 gets executed 1000 times. Since at every evaluation of the loop condition the symbolic executor will need to consider two cases and fork a new path to handle one of the cases, 1000 paths will be generated before the {\tt if} 
condition at line 4 evaluates to true. Let's assume that we are interested in 
knowing whether the {\tt foo} function can be reached or not. 
Considering the fact that {\tt Y} is also a symbolic value, targeting the analysis to 
{\tt foo} has the benefit to eliminate the 1000 paths that will take the else branch 
and focus on the 1000 paths that will reach {\tt foo}.


The example in Figure \ref{fig:needforguidance} also includes a NULL dereferencing error involving 
the pointer {\tt p} at line 8, which is 
due to the last definition of {\tt p} assigning a NULL value at line 6. 
Let's assume that line 8 is set as the target. 
A targeted analyzer that moves backward starting from the target can detect the bug by just focusing on the part of the path defined by lines 6-8. 
There are two challenges to be dealt by the backward executing  analyzer.
The first challenge in backward execution is that statements will need to be evaluated without knowing which definitions or {\em def}s, they should use as the defining statement has not 
been executed yet. Our high-level solution is to introduce an abstract 
address space and record the side effect of the statements as expressions defined over the abstract objects, which will get 
resolved later using a concrete address space. Note that our goal is not to formally verify the program. Instead, our goal is to detect bugs precisely within a specific part of the program and to generate 
test cases that can reach the targets. Therefore, our backward execution approach does not intend to compute the weakest precondition. Our approach is designed to perform path-precise analysis, namely dynamic symbolic execution, in backward mode. 

The second challenge is claiming the reachability of the error, which requires 
dealing with the the path explosion due to line 5. Our solution is to have the backward execution to perform a bounded analysis, e.g., generating paths that execute the loop body up to a specific number of loop iterations, and to explore a small subset of the candidate paths. 
Although this may not allow the backward executor to claim the reachability of the target, 
it allows the collection of constraints that can guide a regular symbolic executor towards the target. 
For instance, through infeasible paths that execute the while loop at line 4
 for a small number of iterations, the targeted analyzer will still generate the constraint {\tt W == 1000} although it will find it to be unsatisfiable when considering other constraints that correspond to the loop condition. 
Such constraints can be solved independently and the models or concrete values, e.g., {\tt 1000}, can be used to guide the forward symbolic executor to the target. 
In this example, after the backward executor infers 1000 for {\tt W}, the forward executor will start its exploration and it will be configured to always return 1000 in response to any read access for {\tt W} that depends on the {\em def} at line 2, causing a 1000X reduction in the number of paths generated before reaching the target.

%% file: approach.tex
\section{Approach}
\label{sec:approach}

In this section, we present the technical details of our Targeted Dynamic Symbolic Execution approach, which consists of a Multi-Pass Backward Dynamic Symbolic Execution (MPBSE) that is followed by an optional Guided Forward Dynamic Symbolic Execution (GFSE).  
Section \ref{sec:notations} presents the notations and definitions.
Section \ref{sec:tse} presents the approach at a high-level. 
Section \ref{sec:mpbse} delves into 
the technical details of MPBSE, specifically, the backward and forward passes in Sections 
\ref{sec:backpass} and \ref{sec:forpass}, respectively. 
Section \ref{sec:replayex} demonstrates the approach on an example.
Section \ref{sec:soundness} discusses soundness of the approach.
Section \ref{sec:implementation} provides some information about the implementation.

\begin{figure}[th!]
\begin{footnotesize}
\begin{equation}
\begin{split}
    \textit{Statement} &::=  \textit{Assignment} \ | \ \textbf{free} \ \textit{Exp}; \ | \ \textit{Statement} \ ; \ \textit{Statement} \\
    & \ | \ \textbf{If} \ (\textit{Cond}) \  \{ \ \textit{Statement} \ \} \ \textbf{Else}  \  \{ \ \textit{Statement} \ \} \ | \ \\ 
 & \ \ \ \ \ \ \ \textbf{While} \ (\textit{Cond}) \ \{ \ \textit{Statement} \ \}  \ | \ \textbf{abort} \\
    \textit{Cond} &::= \textit{Exp} \ | \ ! \textit{Exp} \ | \ \textit{Exp} \ \textit{Relop} \ \textit{Exp}  \\   
    \textit{Relop} &::= \ > \ | \ < \ | \ >= \ | \ <= \ | \ = \ | \ != \\ 
    \textit{Assignment} &::=  \textit{Exp} = \textbf{newSymbolic}  \ | \  
     \textit{Exp} = \textit{Exp} \ | \ \\
    & \textit{Exp} =  \textbf{malloc} \ \textit{Constant} \\  
    \textit{Exp} &::= p \ | \ x \ | \ p \rightarrow f \ | \ p[\textit{Exp}] 
     \ | \ *p \ | \ \&p \\
        &  \textit{Exp} \ \textit{AOp} \ \textit{Exp} \ | \ - \textit{Exp} \ | \
        \ \textit{Constant}\\
     \textit{Aop} &::= + \ | \ - \ | \  \%  \ | \  /  \ | \  * \\      
\end{split}    
\end{equation}
\end{footnotesize}
\caption{Syntax for a simple while language with pointers.}
\label{fig:syntaxwhile}
\end{figure}

\begin{table}[th!]
\centering
\caption{The components of $B=(\mathcal{O},\mathcal{S},\Phi,\mathcal{U},\mathcal{E},\textit{pc})$.}
\label{tab:notations}
\begin{tabular}{c|r} \toprule  
   {\bf Notation} & {\bf Definition} \\ \hline 
      $\mathcal{S}$    & Set of symbolic values \\
   $\mathcal{A}$  &  Set of abstract objects\\
   $\mathcal{H}$   & Set of heap objects \\
   $\mathcal{G}$   & Set of global objects \\ 
   $\mathcal{O}$ &  $\mathcal{A} \cup \mathcal{H} \cup \mathcal{G}$ \\ 
   $\Phi: \mathcal{O} \times \mathcal{N} \mapsto \mathcal{O} \times \mathcal{N}$ & Points-to map \\
   $\textit{u}: \mathcal{O} \times \mathcal{N} \times \mathcal{O} \times \mathcal{N}$ & Unification pair\\
   $\mathcal{U}: \mathcal{P}(\mathcal{O} \times \mathcal{N} \times \mathcal{O} \times \mathcal{N} )$ & Set of Unification pairs \\
   $\textit{w}: \mathcal{O} \times \mathcal{N} \mapsto Expr $ & Store Tuple \\
   $\textit{c}: Expr$ & Constraint \\
   $\delta: (\textit{u},\textit{w},\textit{c})$ & Side effect of executed instruction $\textit{pc}$ \\
   $\mathcal{E}: \textit{List of } \delta$ & Side effect of executed instructions \\
   & (execution history)\\
   $\textit{pc}$ & Last instruction executed\\ 
   & in backward mode \\
   \bottomrule
\end{tabular} 
\end{table}

\subsection{Notations}
\label{sec:notations}

In this section, we present the notations we use to describe the technical approach. 
To simplify the discussion, we define the semantics of the approach using a simplified {\tt while} language with pointers and arrays as shown in Figure \ref{fig:syntaxwhile} (see Section \ref{sec:implementation} for details about an implementation over LLVM for inter-procedural analysis).

Table \ref{tab:notations} presents the definitions of symbols for defining the components of a backward symbolic execution state (or path) $B=(\mathcal{O},\mathcal{S},\Phi,\mathcal{U},\mathcal{E},\textit{pc})$. 
The set of abstract objects is denoted by $\mathcal{A}$ while the concrete address space 
consists of the set of heap objects, $\mathcal{H}$, and the set of global objects,  $\mathcal{G}$\footnote{Our implementation also keeps track of local objects along with the relevant activation frames for the inter-procedural analysis.}. 
$\Phi$ represents the points-to function that maps a pointer in the form 
of a pointer object and a byte offset, $(o_1,f_1)$ to a memory location in the form of the pointed object and a byte offset, $(o_2,f_2)$, i.e., $R(o_1,f_1)=\&o_2 + f_1$, where $R$ represents the read operation.
$\mathcal{O}$ represents the union of abstract and concrete objects. 
$\delta$ represents the side effect of the most recently executed statement   $\textit{pc}$, as a 
combination of $u$, $w$, and $c$,   
and $\mathcal{E}$ represents the sequence of executed statements in the form of a 
list of $\delta$s, where $\mathcal{E}[0]$ represents the side effect of the target statement or the first statement executed in backward direction.
$u$ represents the unification tuple, if any.
$\mathcal{U}$ represents the pairs of unified objects along with their respective offsets. $\textit{w}$ represents a store operation by a combination of the target location, which is defined as a combination of an object and an offset into it, and 
the expression for the stored value. 
$c$ represents the branch condition asserted based on $\textit{pc}$.
$\bot$ is used to represent undefined values.
We use $p$, $x$, $a$, $v$, $pexp$ represent a pointer type variable, a primitive type variable, an array, a value, and a pointer expression, respectively.
We use $\textit{ao}_{\textit{new}}$, $\textit{ho}_{\textit{new}}$, $s_{\textit{new}}$ to denote a newly created abstract object, heap object, and a symbolic value, respectively.

\begin{table}[th!]
    \centering
    \caption{The definition of $A(B,\textit{Expr})$ for different cases.}
    \label{table:addressexp}
    \begin{tabular}{c|c} \toprule
        $\textit{Expr}$ & $A(B,\textit{Expr})$ \\ \hline
        Constant, NULL &  $(\bot,0)$ \\
        $v$,$p$,$a$,$\&p$ &  $(v,0)$,$(p,0)$,$(a,0)$,$(p,0)$ \\
        $*p$ &  $B.\Phi(p,0)$ \\
        $a[exp]$ & $(a,ElementSize(a)*exp)$ \\ 
        $pexp \rightarrow f$ & $(B.\Phi(A(B,pexp)) \oplus \textit{Offset}(o,\textit{f})) $\\
        $pexp[exp]$ & $(A(B,pexp) \oplus ElementSize(pexp)* exp )$ \\ 
        $\&pexp$ & $A(B,pexp)$\\
        $*pexp$ & $B.\Phi(A(B,pexp))$ \\
        \bottomrule
    \end{tabular}
\end{table}

Function $A(B,\textit{Expr})$, the address expression function, maps an expression $\textit{Expr}$ in the context of a backward symbolic execution state $B$ to 
an address expression $\textit{ae}$, which we describe with an object $o$ and a byte offset $f$, i.e., $\textit{ae}=(o,f)$, and 
the individual components of the address expression $\textit{ae}$ can be accessed as $o=\textit{ae}.\textit{object}$ and 
$f=\textit{ae}.\textit{offset}$. 
We use $\textit{ae} \oplus \textit{ov}$ as a short hand notation for 
$(\textit{ae}.\textit{object},\textit{ae}.\textit{offset} + \textit{ov})$, where 
$\textit{ov}$ can be any numeric expression.
Table \ref{table:addressexp} shows how $A(B,\textit{Expr})$ is defined 
for different cases of $\textit{Expr}$. When it is obvious from the context, 
we will drop $B$ and use the notation $A($\textit{Expr}$)$.
Finally, we use the notation $e[\textit{old}/\textit{new}]$ to denote replacing every occurrence of $\textit{old}$ by $\textit{new}$ in $e$.

\subsection{Targeted Symbolic Execution (TSE)}
\label{sec:tse}

{\em TSE} aims to deal with the path explosion problem by starting the analysis from a 
program statement of interest, which we call the {\em target}, and 
working backwards until it reaches the entry point while handling pointers, arrays, and arbitrary pointer arithmetic in a precise way. 
Algorithm \ref{alg:TSE} shows how TSE works. It starts with a Multi-Pass Backward Dynamic Symbolic execution (MPBSE) stage with respect to a target and generates a mix of feasible and infeasible paths that conforms to the program's control-flow. 
If it finds feasible paths, i.e., with satisfiable path conditions, then it returns 
those along with any errors encountered.
However, since the path explosion problem may also be an issue for backward analysis, 
MPBSE may not be able to find any feasible paths. 
TSE can be configured to have a standard (forward) symbolic execution (FSE) step executed 
after MPBSE. The idea is to have the FSE step leverage the values inferred for some of the symbolic values in the MPBSE stage and got stored in what we call a $\textit{replayMap}$, to avoid path explosion in reaching the target. 

\begin{algorithm}[th!]
\begin{footnotesize}
\caption{The algorithm for Targeted Symbolic Execution.}
\label{alg:TSE}
\begin{algorithmic}[1]
\State {\bf TargetedSymbolicExecution}($P$: Program, $E$: Stop, $\textit{target}$: Code Location, $\textit{tuningPar}$: MPBSE Tuning Parameters,
$\textit{fseStep}: \{\textit{True},\textit{False}\}$)
\State {\bf Output} $(\textit{reachable},\textit{errors})$
\State $(\textit{reachable},\textit{errors},\textit{replayMap}) \gets \ $  \textit{MPBSE}($P$,$E$,$\textit{target}$,$\textit{tuningPar}$)
\If{$\neg \textit{reachable}$ and $\textit{fseStep}$} \Comment{GFSE}
   \State Let $s_{\textit{init}}$ denote the initial state of $P$
   \State $\textit{fstates} \gets \{s_{\textit{init}}\}$
   \While{$\textit{fstates} \not = \emptyset$}
      \State $\textit{fs} \gets \textit{removeOne}(\textit{fstates})$
      \If{$\textit{fs}.\textit{pc} = \textit{target}$}
         $\textit{reachable} \gets true$ \ ; \ {\bf break}
      \EndIf
      \If{$\textit{fs}.\textit{pc} \not \equiv lhs = \textit{newSymbolic}$ or $\textit{replayMap}(\textit{fs}.\textit{pc}) = \bot$}
          \State $\textit{preds} \gets \textit{StandardExecuteInstruction}(\textit{fs},\textit{fs}.\textit{pc},\textit{errors})$
         \Else
            \State Let $(o,f)$ denote the object offset pair for $\textit{lhs}$
            \State $\textit{write}(\textit{fs},(f,Sizeof(lhs)),\textit{replayMap}(\textit{st}))$ 
            \State $\textit{preds} \gets \textit{preds} \cup \{fs\}$
         \EndIf
         \State $\textit{fstates} \gets \textit{fstates} \cup \textit{preds}$
   \EndWhile
\EndIf
\State {\bf return} $(\textit{reachable},\textit{errors})$   
\end{algorithmic}
\end{footnotesize}
\end{algorithm}

\subsection{Multi-Pass Backward Symbolic Execution (MPBSE)}
\label{sec:mpbse}

MPBSE consists of a backward pass (Section \ref{sec:backpass}) that uses an abstract address space to speculate 
about the side effect of individual statements on the abstract address space and a forward pass (Section \ref{sec:forpass}) that unifies the 
abstract and concrete objects, if possible, and propagates them in the forward direction to check for errors and to check for the feasibility of the generated paths.

\begin{algorithm}[th!]
\begin{footnotesize}
\caption{The algorithm for Multi-Pass Backward Symbolic Execution (MPBSE).}
\label{alg:MPBSE}
\begin{algorithmic}[1]
\State {\bf MPBSE}($P$: Program, $E$: Stop, $\textit{target}$: Code Location, $\tau_{\textit{edge}}$, $\tau_{\textit{fork}}$: Structural Thresholds, $\textit{evalScheme}$: \{unificationTime, periodic\})
\State {\bf Output} $(\textit{reachable},\textit{errors}, $\textit{replayMap}$)$
\State Let $g_{\textit{init}} \gets \lambda g. \textit{InitialValue}(P,g)$
\State $\textit{replayMap} \gets \lambda \textit{st}. \bot$ \ ; \  $\textit{reachable} \gets \textit{False}$ \ ; \ $\textit{errors} \gets \emptyset$
\State Let $s_{\textit{init}} \gets (P.\textit{Globals},\emptyset,\lambda o.f.\bot,\emptyset,[],\textit{target})$
\State $\textit{states} \gets \{s_{\textit{init}}\}$
\While{$\textit{states} \not = \emptyset$}
   \State $s \gets \textit{removeOne}(\textit{states})$
   \If{$s.\mathcal{E}.\textit{length} \geq \tau_{\textit{path}}$}
      \ {\bf continue}
   \EndIf
   \State $\textit{preds} \gets \textit{ExecuteBackward}(P,s,\tau_{\textit{edge}},\tau_{\textit{fork}})$
   \State $\textit{states} \gets \textit{states} \cup \{\textit{preds}\}$
    \For{each $\textit{ps} \in \textit{preds}$}
         \If{time to do a forward pass based on $\textit{evalScheme}$}
            \For{$i:B.\mathcal{E}.length - 1 \ to \ 0$}
                \If{$B.\mathcal{E}[i].u = ((e_1,f_1),(e_2,f_2))$}
                   \State Let $\mathcal{R}(x)$ denote  $Resolve(B,x,g_{\textit{init}},i)$ 
                   \State $u' \gets ((\mathcal{R}(e_1),\mathcal{R}(f_1)),(\mathcal{R}(e_2),\mathcal{R}(f_2)))$
                   \If{$u \not = u'$}
                     \ $B.\mathcal{E}[i].u \gets u'$ \ ; \
                      \ $B.\mathcal{U} \gets B.\mathcal{U}[u/u']\}$
                   \EndIf
                 \EndIf  
       \EndFor
       \State $ps.\mathcal{U} \gets \textit{TransitiveClosure}(ps.\mathcal{U})$
       \State $\textit{resolved} \gets \emptyset$
       \For{each $0 \leq i < B.\mathcal{E}.\textit{length}$ s.t. $ B.\mathcal{E}[i].\delta.c \not = \textit{True}$}
          \If{$\textit{Resolve}(\textit{ps},B.\mathcal{E}[i].\delta.c,g_{\textit{init}},\textit{i}-1,(B.pc=E))$}
            \State $\textit{resolved} \gets \textit{resolved} \cup B.\mathcal{E}[i].\delta.c$
          \EndIf
       \EndFor
       \If{$\textit{f} \gets \textit{isSAT}(\bigwedge_{r \in \textit{resolved}} r)$}
          \If{$\textit{ps}.\textit{pc}=E$} \Comment{The state has terminated }
              \State $\textit{reachable} \gets \textit{true}$
              \State $\textit{errors} \gets \textit{errors} \cup \textit{checkAndFindErrors}(ps)$
               \State $\textit{states} \gets \textit{states} \setminus \{ps\}$
           \EndIf
       \Else \Comment{Collect constraints for guidance}
           \State $\textit{states} \gets \textit{states} \setminus \{ps\}$
            \State Let $\textit{resolved'} \gets \textit{SortBackwardExecutionOrder}(\textit{resolved})$
            \For{each $i: 1 \ to \ \textit{resolved'}.\textit{length}$}
                \State Let $\textit{symValuePairs} \gets Solve(\textit{resolved'}[i])$
                \For{each $(\textit{sym},\textit{value}) \in \textit{symValuePairs}$}
                   \State Let $\textit{st}$ denote the statement generating $\textit{sym}$ 
                   \If{$\textit{replayMap}(\textit{st})=\bot$}
                      \State $\textit{replayMap} \gets \textit{replayMap}[\textit{st} \mapsto \textit{value}]$
                   \EndIf
                \EndFor
            \EndFor
        \EndIf  
       \EndIf
    \EndFor
\EndWhile
\State {\bf return} $(\textit{reachable},\textit{errors},\textit{replaySet})$
\end{algorithmic}
\end{footnotesize}
\end{algorithm}

Algorithm \ref{alg:MPBSE} shows how the two passes are combined to determine 
if the target is reachable, ($\textit{reachable}$), to find the set of erroneous feasible states, ($\textit{errors}$), and to infer the values that can be replayed by a forward symbolic execution to reach the target ($\textit{replayMap}$). 
First the initial state gets created at line 5 by setting the program counter $\textit{pc}$ to the target and initializing the address space with the global variables in the program $P$. The set of states to explore is initialized to include this initial state at line 6 and while there are states to explore (lines 7-45), in each iteration it chooses a 
state to explore in backward mode by executing the next statement backward and including all the predecessors in the set of states to explore (lines 11-12).
Since a forward pass is expensive, MPBSE can be configured by an evaluation scheme $\textit{evalScheme}$ that determines how frequently it should be performed (line 14). 
The options are 1) immediately after executing every statement that performs unification 
($\textit{unificationTime}$) and 2) after every certain number of statements executed ($\textit{periodic}$).

\subsubsection{Backward Pass}
\label{sec:backpass}

\begin{algorithm}[th!]
\caption{The algorithm for executing a statement backward. $\textbf{PEB}$ is defined in Figure \ref{fig:sem1}.}
\label{alg:execback}
\begin{footnotesize}
\begin{algorithmic}[1]
\State{\bf ExecuteBackward}($P$: Program, $B$: Backward Execution State, $\tau_{\textit{edge}}$: Edge Limit, $\tau_{\textit{fork}}$: Fork Limit)
\State {\bf Output} $\textit{preds}$: Set of Backward Execution States 
\State $\textit{preds,predpc} \gets \emptyset$
\If{$B.\mathcal{E} = []$}
  $\textit{predpc} \gets \{B.pc\}$
\Else
  $\ \textit{predpc} \gets \textit{Predecessor}(P.\textit{CFG},B.pc)$
\EndIf
\For{$\textit{st} \in  predpc$ s.t. $\tau_{\textit{edge}}$ and $\tau_{\textit{fork}}$ not exceeded} \Comment{Possible forking}
  \State Let $B' \gets B$
 \If{$\textit{st}$ is a control-flow statement}
   \If{$\textit{st} \equiv abort$} 
      \If{$B'.\mathcal{E} \not = []$}
         {\bf continue} 
      \EndIf
   \ElsIf{$\textit{st} \equiv if \ e$}
         $B'.(\mathcal{O},\Phi,\mathcal{U}) \gets \textbf{PEB}(B',e)$ \ ; \ $B'.\delta.c = e$
   \ElsIf{$\textit{st} \equiv else$ of $if \ e$}
        \State $B'.(\mathcal{O},\Phi,\mathcal{U}) \gets \textbf{PEB}(B',e)$ \ ; \ $B'.\delta.c = \neg e$
   \ElsIf{$\textit{st} \equiv while \ e$}
      \State  $B'.(\mathcal{O},\Phi,\mathcal{U}) = \textbf{PEB}(B',e)$ 
      \If{$B'.ps$ is covered by $\textit{st}$}
          $B'.\delta.c = e$
       \Else \ 
         $B'.\delta.c = \neg e$
       \EndIf
   \EndIf
 \Else
    \If{$\textit{st}  \equiv lhs = newSymbolic$}
        \State $B'.\mathcal{S} = B'.\mathcal{S} \cup \{s_{\textit{new}}\}$
        \State $B'.\delta=(\bot,(\textbf{A}(B',lhs),s_{\textit{new}}),\textit{true})$
    \ElsIf{$\textit{st}  \equiv lhs = \textit{rhs}$ and $\neg \textit{isPointerType}(\textit{lhs})$}
         \State $B'.(\mathcal{O},\Phi,\mathcal{U}) \gets \textbf{PEB}(B',\textit{rhs})$
         \State $B'.\delta=(\bot,(\textbf{A}(B',lhs),rhs)),\textit{true})$
    \Else 
    \If{$\textit{st} \equiv lhs = malloc \ C$}
        \State $B'.(\mathcal{O},\Phi,\mathcal{U}) \gets \textbf{PEB}(B',\textit{st})$ 
       \State Let $B'.\mathcal{O}=B'.\mathcal{O} \cup \{\textit{ho}_{\textit{new}}\}$
       \State $B'.\delta=(\bot,(\textbf{A}(B',lhs),\&\textit{ho}_{\textit{new}}),\textit{true})$ 
    \ElsIf{$\textit{st} \equiv lhs = rhs$}
        \State $B'.(\mathcal{O},\Phi,\mathcal{U}) \gets \textbf{PEB}(B',\textit{rhs})$
        \State $B'.(\mathcal{O},\Phi,\mathcal{U}) \gets \textbf{PEB}(B',\textit{st})$ 
       \State $B'.\delta=(\bot,(\textbf{A}(B',lhs),rhs),\textit{true})$  
    \Else   
       $B'.\delta=(\bot,\bot,\textit{true})$
    \EndIf
    \EndIf   
  \EndIf
  \State $B'.\textit{pc}=\textit{st}$ \ ; \ $add(B'.\mathcal{E}, B'.\delta)$ \ ; \
  $\textit{preds} \gets \textit{preds} \cup \{B'\}$
\EndFor
\State {\bf return} $\textit{preds}$
\end{algorithmic}
\end{footnotesize}
\end{algorithm}

The role of the backward pass is to interpret the side effect of each 
program statement within the context of the current state $B$ according to the program semantics by creating expressions using 
an abstract address space and to create a backward search tree to explore
different execution paths that are valid according to the control-flow. 
We will provide an example of the backward pass later using the sample code 
in Figure \ref{fig:needprecision} and the step-by-step evaluation in Table \ref{table:eval}, but we will first introduce the details using  Algorithm \ref{alg:execback} and Figure \ref{fig:sem1}.

Algorithm \ref{alg:execback} determines the next statement(s) to execute 
using the control-flow graph of the program (line 5) unless no statement 
has been executed yet, in which case the target statement is used (line 4).
For each statement to be executed (line 7) it creates a new state $B'$ 
by initializing it from the current state $B$ (line 8).
If the next statement to execute is a control-flow statement (lines 9-21), either the path may get terminated (for {\tt abort}) (lines 10-12) , or the relevant constraint gets recorded 
(lines 13-21). 
If the next statement to execute is an assignment, different cases are considered. 
Non-pointer assignments are handled at lines 23-28. 
Pointer related aspects of statements and expressions are handled according to the operational semantics defined 
in Figure \ref{fig:sem1}, which we denote by $\textit{PEB}$ or $\textit{PointerExecuteBackward}$. 

The rules in Figure \ref{fig:sem1} explain how the set of objects in the whole address space $\mathcal{O}$, 
the set of unification pairs $\mathcal{U}$, and the points-to map $\Phi$ 
gets updated for different cases of expressions (E) and statements (S). 
A common theme in  these rules is that 
if the right-hand side expression $\textit{rhs}$ in an expression or an assignment does not 
point-to anything ($\bot$) based on the backward execution so far then a 
new abstract object $\textit{ao}_{\textit{new}}$ gets created and added to 
$\mathcal{O}$ and the points-to map is updated in the next backward execution state 
for $\textit{rhs}$ to be mapped to $\textit{ao}_{\textit{new}}$. 
If the left-hand side expression $\textit{lhs}$ of an assignment does not 
point-to anything ($\bot$) then no unification happens. 
Otherwise, the object pointed by $\textit{lhs}$ gets unified with 
either the newly introduced abstract object for $\textit{rhs}$ or 
what $\textit{rhs}$ is known to be pointing-to $\Phi({\textit{rhs}})$.
Also, if $\textit{lhs}$ is known to be pointing to some object in the current 
state then it gets cleared (or killed) in the next backward execution state by setting it to $\bot$ 
because the statement overwrites that points-to relationship in forward semantics. 
However, as new {\em use}s or right-hand side appearances of that $\textit{lhs}$ gets encountered in backward mode, a new abstract object will be created to refer to it. 
These steps guarantee that abstract objects get unified with either other abstract objects or some concrete objects according to the standard {\em def}-{\em use} relationship in program reasoning.

\begin{figure}[th!]
\begin{footnotesize}
\begin{Verbatim}[numbers=left,xleftmargin=5mm]
        struct A {int f1; int f2;} p,q;
        q = malloc(sizeof(struct A));
        p = q;
        q->f1 = 5;
        p->f1 = 7;      
        if (p->f1 == q->f1) 
        fail:   assert(0); 
}
\end{Verbatim}
\end{footnotesize}
\caption{An example that involves pointers and aliasing.}
\label{fig:needprecision}
\end{figure}

So, in Algorithm \ref{alg:execback}, $\textit{PEB}$ gets called to 
handle pointer related side effects of expressions and statements  
 according to Figure \ref{fig:sem1}. 
The remaining side-effects, mainly the store information, gets computed 
based on the specific cases. 
As an example consider the sample code in Figure \ref{fig:needprecision}, where 
the target statement is the {\tt assert} statement at line 7. 
Table \ref{table:eval} shows the step-by-step affect of executing 
each statement in backward mode.
So, executing backward starting from line 7, first  
{\tt p->f1} and {\tt q->f1} will be evaluated as right-hand side expressions. 
Since the points-to relation maps these variables to $\bot$ at step 1,
abstract objects $a_1$ and $a_2$ of type {\tt struct A} get created 
and {\tt q} and {\tt p} are made 
to point to these abstract objects after steps 2 and 3, respectively.
Evaluation of the {\tt if} statement at line 6 is performed by reading from 
these abstract objects and recording the constraint $R(a1,F1) = R(a2,F1)$, where $R$ 
and $F1$ denote the read function and the offset of the  {\tt f1} field in {\tt struct A}.
The side-effect of executing {\tt p->f1=7} backward will be recording the 
store information in $\delta.w$ as $((a_2,F1),7)$. 
Similarly, the side effect of {\tt q->f1=5} will be recorded as 
$((a_1,F1),5)$ in $\delta.w$.
There will be three side effects of executing the next statement, $p = q$. 
The first one is the unification of $a_1$ and $a_2$, which is recorded as $((a_1,0),(a_2,0))$. 
The second one is the store action, which is recorded as $((p,0),\&\Phi(q,0))$, 
which would evaluate to $((p,0),\&(a_1,0))$.
The third one is the kill action regarding the points-to value for $p$.
Finally, executing the statement  {\tt q = malloc(...)} in backward mode would introduce a 
new heap object $h_1$ that gets unified with $a_1$, which is recorded as 
$((h_1,0),(a_1,0))$. The other side effects include resetting the points-to value for $q$ 
and recording the store information as $((q,0),\&(h_1,0))$.

\begin{figure*}[th!]
\begin{footnotesize}
\begin{equation*}
\begin{split}
[E1: \ \textit{rhs}] \ \frac{
  \ \ B'.(\mathcal{O},\Phi,\mathcal{U})=\textbf{PEB}(B) \
\ \ (B.\Phi(\textbf{A}(rhs)) \not = \bot \ \vee \neg \textit{isPointer}(\textit{rhs})) 
}
{
  B'.(\mathcal{O},\Phi,\mathcal{U}) = B.(\mathcal{O},\Phi,\mathcal{U})
}
\end{split}
\end{equation*}
\begin{equation*}
\begin{split}
[E2: \ \textit{rhs}] \ \frac{
  \ \ B'.(\mathcal{O},\Phi,\mathcal{U})=\textbf{PEB}(B) \
\ \ B.\Phi(\textbf{A}(rhs)) = \bot \ \textit{isPointer}(rhs)
}
{
B'.\mathcal{O}= B.\mathcal{O} \cup \{\textit{ao}_{\textit{new}}\} \ B'.\Phi=B.\Phi[\textbf{A}(rhs) \mapsto (\textit{ao}_{\textit{new}},0)] \ B'.\mathcal{U}=B.\mathcal{U}  \ 
}
\end{split}
\end{equation*}
\begin{equation*}
\begin{split}
[E3: \ \textit{Uop} \ \textit{rhs}] \ \frac{
  \ \ B'.(\mathcal{O},\Phi,\mathcal{U})=\textbf{PEB}(B) 
}
{
B'=\textbf{PEB}(B,\textit{rhs}) 
}
,
[E4: \ \textit{rhs}_1 \ \textit{Bop} \ \textit{rhs}_2] \ \frac{
  \ \ B'.(\mathcal{O},\Phi,\mathcal{U})=\textbf{PEB}(B) 
}
{
B'=\textbf{PEB}(\textbf{PEB}(B,\textit{rhs}_1),\textit{rhs}_2) 
}
\end{split}
\end{equation*}
\begin{equation*}
\begin{split}
[S1: \ lhs = rhs] \ \frac{
  \ \ B'.(\mathcal{O},\Phi,\mathcal{U})=\textbf{PEB}(B) \
\ \ B.\Phi(\textbf{A}(lhs)) = \bot \ B.\Phi(\textbf{A}(rhs)) = \bot
}
{
B'.\mathcal{O}= B.\mathcal{O} \cup \{\textit{ao}_{\textit{new}}\} \ B'.\Phi=B.\Phi[\textbf{A}(rhs) \mapsto (\textit{ao}_{\textit{new}},0)] \ B'.\mathcal{U}=B.\mathcal{U}  \ 
}
\end{split}
\end{equation*}
\begin{equation*}
\begin{split}
[S2: \ lhs = rhs] \ \frac{
  \ \ B'.(\mathcal{O},\Phi,\mathcal{U})=\textbf{PEB}(B) \
\ \ B.\Phi(\textbf{A}(lhs)) \not = \bot \ B.\Phi(\textbf{A}(rhs)) = \bot
}
{
B'.\mathcal{O}= B.\mathcal{O} \cup \{\textit{ao}_{\textit{new}}\} \ B'.\Phi=B.\Phi[\textbf{A}(rhs) \mapsto (\textit{ao}_{\textit{new}},0)][\textbf{A}(lhs) \mapsto \bot] \ B'.\mathcal{U}=B.\mathcal{U} \cup \{(\Phi(\textbf{A}(lhs)),(\textit{ao}_{\textit{new}},0))\} 
}
\end{split}
\end{equation*}
\begin{equation*}
\begin{split}
[S3: \ lhs = rhs] \ \frac{
  \ \ B'.(\mathcal{O},\Phi,\mathcal{U})= \textbf{PEB}(B) \
\ \ B.\Phi(\textbf{A}(lhs))  = \bot \ B.\Phi(\textbf{A}(rhs)) \not = \bot
}
{
  B'.(\mathcal{O},\Phi,\mathcal{U}) = B.(\mathcal{O},\Phi,\mathcal{U})
}
\end{split}
\end{equation*}
\begin{equation*}
\begin{split}
[S4: \ lhs = rhs] \ \frac{
  \ \ B'.(\mathcal{O},\Phi,\mathcal{U})=\textbf{PEB}(B) \
\ \ B.\Phi(\textbf{A}(lhs)) \not = \bot \ B.\Phi(\textbf{A}(rhs)) \not = \bot
}
{
B'.\mathcal{O}= B.\mathcal{O}  \ B'.\Phi=B.\Phi[\textbf{A}(lhs) \mapsto \bot] \ B'.\mathcal{U}=B.\mathcal{U} \cup \{(\Phi(\textbf{A}(lhs)),\Phi(\textbf{A}(rhs))\} 
}
\end{split}
\end{equation*}
\begin{equation*}
\begin{split}
[S5: \ lhs = malloc \ C] \ \frac{
\ \  B'.(\mathcal{O},\Phi,\mathcal{U})=\textbf{PEB}(B)  \ B.\Phi(\textbf{A}(lhs)) = \bot
}
{
  B'.(\mathcal{O},\Phi,\mathcal{U}) = B.(\mathcal{O},\Phi,\mathcal{U})
}
\end{split}
\end{equation*}
\begin{equation*}
\begin{split}
[S6: \ lhs = malloc \ C] \ \frac{
\ \  B'.(\mathcal{O},\Phi,\mathcal{U})=\textbf{PEB}(B)  \ B.\Phi(\textbf{A}(lhs)) \not = \bot
}
{
B'.\mathcal{O}= B.\mathcal{O} \cup \{\textit{ho}_{new}\} \ B'.\Phi=B.\Phi[\textbf{A}(lhs) \mapsto \bot] \ B'.\mathcal{U}=B.\mathcal{U} \cup \{(\textit{ho}_{new},0),B.\Phi(\textbf{A}(lhs)))\} 
}
\end{split}
\end{equation*}
\begin{equation*}
\begin{split}
[S7: \ free \ rhs] \ \frac{
\ \  B'.(\mathcal{O},\Phi,\mathcal{U}) = \textbf{PEB}(B)  \ B.\Phi(\textbf{A}(rhs)) = \bot
}
{
  B'.(\mathcal{O},\Phi,\mathcal{U}) = B.(\mathcal{O},\Phi,\mathcal{U})
}
,
[S8 : free \ rhs] \ \frac{
\ \  B'.(\mathcal{O},\Phi,\mathcal{U}) = \textbf{PEB}(B)  \ B.\Phi(\textbf{A}(rhs)) \not = \bot
}
{
Error(B')
}
\end{split}
\end{equation*}
\begin{equation*}
\begin{split}
\end{split}
\end{equation*}
\end{footnotesize}
\caption{$\textit{PointerExecuteBackward}$ ($\textit{PEB}$): Backward Symbolic Execution operational semantic with respect to pointer related side effects on $(\mathcal{O},\Phi,\mathcal{U})$. $\textbf{A}(e)$ is a shorthand notation for $A(B,e)$ (see Table \ref{table:addressexp}). $\textit{PEB}(B)$ is a shorthand notation for $\textit{PEB}(B,\textit{CS})$ for the syntax specified as $[E:\textit{CS}]$ or $[S:\textit{CS}]$ for expressions or statements, respectively. $\textit{Uop}$ and $\textit{Bop}$ denote unary and binary arithmetic, relational, or logical operators, respectively.}
\label{fig:sem1}
\end{figure*}

\begin{table*}[th!]
\centering
\caption{The details of a backward pass for the code in Figure \ref{fig:needprecision} with the label {\tt fail}. $F1$ denote the byte offset for field {\tt f1} in {\tt struct A} and $R(o,f)$ denotes read of object $o$ starting from offset $f$. }
\label{table:eval}
\begin{footnotesize}
    \begin{tabular}{c|c|c|c|c|c|c|c} \toprule
       {\bf Program} & {\bf Step} & \multicolumn{6}{c}{\bf Metadata} \\ \cline{3-8}
       {\bf Statement/Exp.} &  & $\mathcal{G}$ & $\mathcal{A}$ & $\mathcal{H}$  & $\Phi$  & $u/w/c$ & $\mathcal{U}$ \\ \hline
        {\tt Initial State} & 1 & $\{p,q\}$ & $\{\}$  & $\{\}$  &  $\{((q,0),\bot),((p,0),\bot)\}$ & - & $\{\}$ \\
       {\tt q->f1} & 2& & & & & &  \\ 
        &  & $\{p,q\}$ & $\{a_1\}$ & $\{\}$ & $\{((q,0),(a_1,0)),((p,0),\bot)\}$& - & $\{\}$  \\
       {\tt p->f1}  & 3 & & & & & &  \\ 
       & & $\{p,q\}$ & $\{a_1,a_2\}$ & $\{\}$ & $\{((q,0),(a_1,0)),((p,0),(a_2,0))\}$ & - & $\{\}$ \\
       {\tt if (p->f1 == q->f1)}  & 4& & & & & &  \\
       & & $\{p,q\}$ & $\{a_1,a_2\}$ & $\{\}$ & $\{((q,0),(a_1,0)),((p,0),(a_2,0))\}$ & $c1=R(a_1,F1) = R(a_2,F1)$ & $\{\}$ \\
       {\tt p->f1 = 7} & 5 & & & & & &  \\
       & & $\{p,q\}$ & $\{a_1,a_2\}$ & $\{\}$ & $\{((q,0),(a_1,0)),((p,0),(a_2,0))\}$ & $w1:((a_2,F1),7)$& $\{\}$ \\
       {\tt q->f1 = 5}  & 6& & & & & &  \\
       & & $\{p,q\}$ & $\{a_1,a_2\}$ & $\{\}$ & $\{((q,0),(a_1,0)),((p,0),(a_2,0))\}$ & $w2:((a_1,F1),5)$& $\{\}$ \\
       {\tt p = q}  & 7 & & & & & &  \\
       & & $\{p,q\}$ & $\{a_1,a_2\}$ & $\{\}$ & $\{((q,0),(a_1,0)),((p,0),\bot)\}$ & $u1:((a_1,0),(a_2,0))$   & $\{u1\}$ \\
       {\tt q = malloc(...)}  & 8& & & & & &  \\
       & & $\{p,q\}$ & $\{a_1,a_2\}$ & $\{\}$ & $\{((q,0),\bot),((p,0),\bot)\}$ & $u2:((h_1,0),(a_1,0))$   & $\{u1,u2\}$ \\
       \bottomrule
    \end{tabular}
    \end{footnotesize}
\end{table*}

\begin{algorithm}[th!]
\caption{The algorithm to resolve an expression.}
\label{alg:resolve}
\begin{footnotesize}
\begin{algorithmic}[1]
\State {\bf Resolve}($B$: BSE State, $e$: Expr, $g_{\textit{init}}: \mathcal{G} \times \mathcal{N} \mapsto \textit{Value}$, $step: \mathcal{Z}$, 
$\textit{errors}$: Set of BSE States, $\textit{terminated}: \{\textit{True},\textit{False}\}$) : $\{\textit{True},\textit{False}\}$
\For{each abstract object $\textit{ao}$ in atomic expression $ae$ in $e$}
   \If{$\exists ((\textit{ao},f_1),(o,f_2))$ s.t. $o$ is not an abstract object}
      \If{$f_2 - f_1 \geq 0$ and $Sizeof(\textit{ao}) - f_1 \leq Sizeof(o) - f_2$}
         \State $ae \gets ae[\textit{ao}/o] \oplus (f_2 - f_1)$
      \Else \ 
       $\textit{errors} \gets \textit{errors}  \cup \{B\}$ \ ; \
       {\bf return} $\textit{False}$
      \EndIf
   \EndIf
\EndFor
\For{each atomic read expression $R(o,f)$ in $e$ s.t. $o$ and $f$ are resolved}
   \If{$\textit{readRange}$ is NOT within the bounds of object $o$} 
       $\textit{errors} \gets \textit{errors}  \cup \{B\}$ \; \ {\bf return false}
   \EndIf
   \State $\textit{readRange} \gets [f,f+\textit{Sizeof}(R(o,f))$
    \If{$\textit{step} < 0$}
       \If{$\textit{terminated}$}
           $e \gets e[R(o,f)/g_{\textit{init}}(o,f)]$
       \Else \ {\bf return false}
       \EndIf
    \Else
     \State $temp \gets 0$
     \For{$i: \textit{step}$ to $|B.\mathcal{E}|-1$ s.t. $B.\mathcal{E}[i].w=((oexp,fexp),vexp)$ }
         \State $dr \gets \textit{Resolve}(B,oexp,g_{\textit{init}},i+1,errors)$   
         \State $or \gets \textit{Resolve}(B,fexp,g_{\textit{init}},i+1,errors)$    
        \State $vr \gets \textit{Resolve}(B,vexp,g_{\textit{init}},i+1,errors)$  
        \If{$\neg dr$ or $\neg or$ or $\neg vr$} {\bf return} $\textit{False}$ 
             \Comment{Cannot determine the def}
        \EndIf    
        \State Let $B.\mathcal{E}[i].w=((o',f'),v)$
        \If{$o=o'$ and $[f',f'+Sizeof(v))$ and $\textit{readRange}$ overlaps}
           \State $\textit{inter} \gets \textit{Intersection}([f',f'+Sizeof(v)),\textit{readRange})$
           \State $\textit{write}(B,\textit{temp}, \textit{inter},v)$
           \State $\textit{readRange} \gets \textit{Remove}(\textit{readRange},\textit{inter})$ 
           \If{$\textit{readRange}$ is empty} 
              \   $e \gets e[R(o,f)/temp]$
              \ {\bf break}
           \EndIf
        \EndIf
     \EndFor
     \If{$\textit{readRange}$ not empty} 
        \If{$\textit{terminated}$}
        \State  $\textit{write}(B,\textit{temp}, \textit{readRange},g_{\textit{init}}(o,f))$
        \State  $e \gets e[R(o,f)/temp]$
        \Else \ {\bf return} $\textit{False}$
     \EndIf
  \EndIf
\EndIf
\EndFor
\State {\bf return} $\neg $($e$ has any unresolved atomic objects or reads)
\end{algorithmic}
\end{footnotesize}
\end{algorithm}

\subsubsection{Forward Pass}
\label{sec:forpass}

The goal of the forward pass is 1) to check the feasibility or infeasibility of the path and 2) to detect errors on the feasible paths.
In Algorithm \ref{alg:MPBSE}, a forward pass is started 
at line 14 for each predecessor state if it satisfies the evaluation scheme $\textit{evalScheme}$.  
As an example, if $\textit{evalScheme}$ is equal to $\textit{unificationTime}$ then
the {\tt if} statement at line 14 would evaluate to true if the backward 
execution of the statement at line 11 has involved the application of the rules 
$S2$, $S4$, or $S6$. 
As an example, when analyzing the code in Figure \ref{fig:needprecision}, 
after step 7 a unification tuple is added to $B.\mathcal{U}$ as shown in 
Table \ref{table:eval}. So, a forward pass can be started 
after step 7. However, this would not be useful in resolving the expressions 
$c1$, $w1$, and $w2$ since only after step 8 
the association with the heap object $h1$ would be discovered.
If $\textit{evalScheme}$ is equal to $\textit{periodic}$ 
then after executing every $p$ statements on each path, the forward pass is enabled, where $p$ is the period value (not shown in the algorithm). 

The first step of the forward pass in Algorithm \ref{alg:MPBSE} is to resolve the 
expressions that appear in the unification tuples (lines 15-22) and then to
compute the transitive closure of 
$B.\mathcal{U}$ while taking the offsets and sizes of the objects into account (line 23). 
For every pair of unification tuples $((o_1,f_1),(o_2,f_2))$ and $((o_3,f_3),(o2,f_4))$ 
such that $f_1$, $f_2$, $f_3$, and $f_4$ are either concrete of symbolic values, the goal is to generate $((o_1,f_5),(o_3,f_6))$ and add it to $B.\mathcal{U}$ by determining $f_5$ and $f_6$. To simplify the 
disposition, we ignore the size information, which is determined by the statement that cause the unification to happen.
If $f_4 \geq f_2$ then $f_6=f_3 - (f_4 - f_2)$ and $f_5=f_1$. Otherwise, $f_5=f_1 - (f_2 - f_4)$ and $f_6=f_3$.

For the sample code in Figure \ref{fig:needprecision}, assuming that the transitive closure is performed after step 8 on $B.\mathcal{U}$ as shown in Table \ref{table:eval}, 
$B.\mathcal{U}$ (or $\textit{ps}.\mathcal{U}$ in Algorithm \ref{alg:MPBSE} at line 15) 
would become $\{((a_1,0),(a_2,0)),((h_1,0),(a_1,0)),((h_1,0),(a_2,0))\}$ after the transitive closure computation and indicates that both $a_1$ and $a_2$ map to the heap object $h_1$ starting at offset 0.

After the transitive closure computation, Algorithm \ref{alg:MPBSE} rewrites the constraints of the path condition, i.e., every $\delta.c$ in the execution history  $B.\mathcal{E}$ by replacing abstract objects and the offsets in the address expressions based on the concrete address expressions they get unified with (lines 24-29) using Algorithm \ref{alg:resolve}, whose goal is to {\em resolve} a given expression within the 
context of a backward execution state.
An expression is deemed as {\em resolved} if it does not involve any abstract objects and all the read expressions that appear in it have been evaluated and replaced with resolved expressions. So, a resolved expression may only have constant and symbolic values, where some of the constant values may refer to the addresses of concrete objects. 

 Algorithm \ref{alg:resolve} takes an expression $e$ and resolves it by rewriting it in the context of a 
 backward execution state $B$ by replacing every abstract object in $e$ with the concrete object that it gets unified with, if any (lines 2-9). 
 So, for the example in Figure \ref{fig:needprecision}, after discovering that 
$a_1$ and $a_2$ map to the heap object $h_1$, it will first rewrite 
$R(a_1,F1) = R(a_2,F1)$ ($c1$) as  $R(h_1,F1) = R(h_1,F1)$, 
$((a_2,F1),7)$ ($w1$) as $((h_1,F1),7)$, and 
$((a_1,F1),5)$ ($w2$) as $((h_1,F1),5)$. Later in this section, we explain how $c1$ gets resolved further.

 After replacing abstract objects and their offsets based on their unification with concrete objects, 
 Algorithm \ref{alg:resolve} evaluates every read expression that involves a concrete object by first 
 resolving the offset and the size expressions of the read operation. If any of these 
 cannot be resolved then it returns false. 
 Our implementation handles symbolic offsets by checking for potential out of bound accesses and by forking a new state for safe and unsafe access cases. 
 However, for simplicity of the discussion, we do not include those details in this algorithm. 
 Assuming that the offset and the size of the read expression have been resolved, 
 it traverses previously executed statements to find the closest or most recent {\em def} 
 that defines the same object on a range that overlaps with the 
 the read range (lines 20-34). If it encounters any {\tt def} whose destination address is not resolved 
then, for soundness, it returns false as it is possible that it may turn out to be a {\em def} of  the same object being read.

Algorithm \ref{alg:resolve} can handle complicated cases where the evaluation of the read expression 
 is defined by multiple {\em def}s as our approach is able to handle pointer arithmetic precisely due to handling of object offsets at the byte granularity.  
 The algorithm uses a temporary object $temp$ with the same size as the read expression to update it in an iterative fashion with the values defined at each relevant {\em def} location
 and rewrite the read expression at one shot (lines 31 and 38).
 We also define a function $\textit{write}(B,o,r,v)$ that writes the value $v$ to the offset range $r$ of object $o$ in state $B$.  
 For that purpose, it updates the range of the read to be evaluated as 
 it analyzes earlier statements (line 31). It completes the search when the read is complete (line 30) or $\textit{readRange}$ becomes empty.
 If the read cannot be completed using the {\tt def} statements that come earlier in the execution order and the state has completed the execution then the value for the remaining range gets computed from the initial state $g\_{\textit{init}}$, which is defined by the initial values of the global and static variables in $P$. 
 The algorithm returns true if after all the rewriting no abstract states 
 or read expressions appear in $e$.
 So, for the example in Figure \ref{fig:needprecision}, the read expression
 $R(h_1,F1)$ in $c1$ in Table \ref{table:eval} will be resolved through the {\em def} {\tt p->f1 = 7}, which was captured in $w1$ as $((h_1,F1),7)$ (see the above explanation on how $a_2$ was replaced by $h_1$). As a result $R(h_1,F1)$ will evaluate to 7 and 
 $c1$ will be rewritten into $7 = 7$, which would later, at line 30 in Algorithm 
 \ref{alg:MPBSE}, be  recognized as a {\tt True} expression when checking the feasibility of the path. So, MPBSE will conclude that the assertion failure at line 7 in Figure \ref{fig:needprecision} is reachable from line 2, the entry point $E$.

In Algorithm \ref{alg:MPBSE}, the resolved constraints are combined to check 
for the feasibility of the state so far using an SMT solver (line 30). If it turns out be satisfiable and the entry 
statement $E$ has been executed then a feasible state that can reach the target 
has been found (line 32). The state is analyzed further for memory errors (see Section \ref{sec:implementation}) (line 33).
Otherwise, the state is not feasible and gets removed from the pool of states (line 34). 
However, the constraints are analyzed independently to find solutions for any symbolic values, which gets returned as an output in $\textit{replayMap}$ (lines 36-48).

\subsection{An Example on Generating Replay Values}
\label{sec:replayex}
In this section, we discuss an example to demonstrate how MPBSE generates the replay values from infeasible paths. 
Consider the code in Figure \ref{fig:needforguidance} and assume that the target statement 
is {\tt r=*p} at line 8, where the program will have a NULL dereferencing error if {\tt W} 
is equal to 1000. However, both DSE and MPBSE would need to generate at least 1000 paths to find this out. 
The following steps will be taken during TSE. First, MPBSE will  
infer that all reads of {\tt W} should use 1000 and return this in $\textit{replayMap}$. 
Then Guided Forward Symbolic Execution will analyze the program, generate a single path by forcing all reads of {\tt W} return 1000, and detect the NULL pointer dereference at line 8 much faster than without the guidance on {\tt W}.

MPBSE will first evaluate the expression $*p$ by mapping $(p,0)$ to an address expression 
$(a_1,0)$ involving a new abstract object 
$a_1$ in $\Phi$. Evaluating $r = *p$ will generate the side effect of the statement as $\delta_1=(\bot,((r,0),R(a_1,0)),\textit{true})$. Then it will evaluate {\tt *p=7} 
by recording the side effect as $\delta_2=(\bot,(\Phi(p,0),7),\textit{true})=(\bot,((a_1,0),7),\textit{true})$. When it evaluates {\tt p=NULL}, it will unify 
$\Phi(NULL,0)=(\bot,0)$ and $\Phi(p,0)=(a_1,0)$ by recording the side effect as 
$\delta_3=(((\bot,0),(a_1,0),((p,0),NULL),\textit{true})$ and by mapping $(p,0)$ to 
$(\bot,0)$. So, no matter what type of statement will be encountered as the execution moves backward there will not be any possibility to unify $a_1$ with any other object. 
Eventually, $a_1$ will be found to be unified with $(\bot,0)$ indicating an 
out of bound access or a NULL pointer dereference. However, due to the path explosion problem, this would require 1000 paths to be generated to determine the feasibility of the path, after which MPBSE can report the memory error. 
Assuming that MPBSE has been configured to limit the number of states to be generated to 1, 
it may generate an infeasible path that does not execute the loop body at line 4  as  follows: 

\begin{footnotesize}
\begin{verbatim}
     i = 0;
     W = newSymbolic;
     !(i<W)
     (W == 1000) 
     p = NULL;
     *p = 7;      
     r = *p;
\end{verbatim}
\end{footnotesize}

After evaluating all the expressions and the statements, the constraints on the path includes $c1:s_1 = 1000$ and $c2:s_1 \leq 0$. In Algorithm \ref{alg:MPBSE} the resolved 
constraints will be sorted to analyze constraints in backward execution order. 
So, $c1$ will be solved before $c2$. Solving $c1$ will map $s_1$ to 1000 and 
map {\tt W = newSymbolic} to 1000 such that any read that uses this {\em def} uses the value 1000. Next it will evaluate $c2$, but since {\tt W = newSymbolic} already has a 
mapping, it will not record the solution for $c2$. The intuition behind this is that 
the constraint closer to the target, e.g., $c1:s_1 = 1000$, will steer the control to the target much better than earlier constraints, e.g., $c2:s_1 \leq 0$.

\subsection{Soundness}
\label{sec:soundness}

\begin{lemma}[Def-Use Correctness]
\label{lem:defuse}
Given a while-program $P$ and a target statement $t$ in $P$, if MPBSE can resolve all 
the constraints in the path condition then every expression in each constraint 
is evaluated using the correct set of defining statements.
\begin{proof}
 This follows from the fact that Algorithm \ref{alg:MPBSE}, Algorithm \ref{alg:execback}, 
 the operational semantic presented in Figure \ref{fig:sem1}, and Algorithm \ref{alg:resolve} computes the points-to information  correctly and precisely, i.e., at the byte granularity, and each byte of every read expression is evaluated using the most recent defining statement that appears before the read expression. 
\end{proof}
\end{lemma}

\begin{theorem}[Soundness]
Given a while-program $P$ and a target statement $t$ in $P$, if MPBSE can resolve all 
the constraints in the path condition then its feasibility decision is equivalent to a byte-precise standard dynamic symbolic execution's decision of $t$'s reachability in $P$.
\begin{proof}
Follows from Lemma \ref{lem:defuse}.     
\end{proof}
\end{theorem}

\subsection{Implementation}
\label{sec:implementation}

We have implemented our approach in a tool called DESTINA using an earlier version of KLEE \cite{CDE08} (KLEE 1.4.0). Our implementation consists of 15K SLOC C++ code and uses LLVM-13. 
DESTINA performs inter-procedural analysis although in this paper we explain the technique over a simple while-language due to space restrictions. 
So, DESTINA supports all the LLVM instructions that are supported in KLEE and represents the 
stack as part of the backward execution state. As in KLEE, it allocates the objects in the concrete address space using {\tt malloc} and evaluates address of symbol (\&) by using the base address of the concrete objects as returned my {\tt malloc}.

Unification is performed locally for each activation frame and then through the unification of abstract objects with the formal and actual arguments and return values,  unification results get propagated to the calling contexts. DESTINA uses the scheduling algorithm and query solving infrastructure provided by KLEE. 

We used two important configuration parameters for DESTINA, which has been referred to as 
MPBSE Tuning Parameters in the input section of Algorithm \ref{alg:TSE} and as 
$\tau_{\textit{edge}}$ and $\tau_{\textit{fork}}$ in the input section of Algorithm \ref{alg:MPBSE}. $\tau_{\textit{edge}}$ is called the in-cycle edge limit and sets an upper bound on the number of times an edge that appears in a cycle can be taken.  $\tau_{\textit{fork}}$ is called the fork limit and sets an upper bound on the number of 
paths to be generated.

\paragraph{Error checking} 
DESTINA performs error checking at two points in the analysis. 
The first one is at the time of resolving read expressions and is achieved by checking whether the range of the read as well as the relevant {\em def}'s ranges do not exceed the object boundaries. The second one is after determining the feasibility of the path. 
DESTINA supports checking out of bound accesses, NULL pointer dereferences, 
double-frees, use-after-frees, and memory leaks. 

%% file: evaluation.tex
\section{Evaluation}
\label{sec:eval}

We evaluated our Targeted Symbolic Execution (TSE) approach by applying it to a set of SvComp benchmarks and comparing its performance with Dynamic Symbolic Execution (DSE) as implemented in KLEE. We ran the experiments three times and averaged the execution times while using median value for other metrics. The experiments have been executed on Ubuntu 22.04 running on WSL2 on a Windows machine with an Intel 11th  Gen i7-1165G7 processor and 32GBs of memory. Our tool DESTINA has been implemented over KLEE 1.4.0. However,
we also extended KLEE 3.1 with memory leak detection capability to include in our evaluation. 

\subsection{\bf RQ1: Can DESTINA Detect Memory Errors?}
\label{sec:rq1}

We used the buggy Svcomp benchmarks in the memory safety category, shown in Table \ref{table:buggymemsafety}, to assess effectiveness of DESTINA in memory error detection. We configured DESTINA's TSE to perform 
only MPBSE to evaluate its bug detection capability independent of DSE. 
We set the {\tt return} statement of 
the {\tt main} function as the target statement for each benchmark. 
We tried different combinations of fork and in-cycle edge limits and Table \ref{table:buggymemsafety} shows the best result for each benchmark. 
DESTINA was able to detect the memory error in 13 out of 18 cases while KLEE 3.1 was able to detect 17 out of 18 within 600 secs. The one benchmark that both of them failed to detect requires more realistic modeling of {\tt malloc}. The ones that DESTINA could not detect either involve loops with a large number of iterations as in {\tt 960521-1-1} 
or consists of many functions with loops or switch statements that can cause path explosion as in {\tt test-0235-3}. In 6 out of 13 cases DESTINA detects the bug much faster than 
KLEE 3.1. {\bf The results show that DESTINA can detect a variety of memory errors very quickly as long as loops with a high number of iterations are not involved on the buggy path.}  

\begin{table*}[th!]
\caption{TSE and DSE on buggy {\tt Svcomp memsafety} benchmarks using a timeout (TO) of 600 minutes.}
\label{table:buggymemsafety}
\centering
\begin{footnotesize}
\begin{tabular}{c|r|c|c|r|r|c|c} \toprule
{\bf Benchmark} & {\bf SLOC} & {\bf Code }& {\bf Bug} &  \multicolumn{2}{c|}{\bf Time (secs)} & \multicolumn{2}{c}{\bf Bug Detected} \\
& & {\bf Features} & {\bf  Type} & {\bf TSE} & {\bf DSE} & {\bf TSE} & {\bf DSE} \\ \hline
{\bf 20020406-1} & 103 & Concrete Loop Condition & Memory Leak & 5.32 & 1.05 & Yes & Yes \\
{\bf 20051113-1} & 72 & Concrete Loop Condition& Memory Leak & 2.16 & 1.02 &  No & Yes\\
{\bf 960521-1-1} & 35 &  Concrete Loop Condition & Invalid Free & - &  1.02 & No & Yes \\
{\bf 960521-1-3} & 28 &  Concrete Loop Condition & Invalid Deref & - & 1.02 & No & Yes\\
{\bf cmp-freed-ptr} & 20 & Address Comparison & Memory Leak & NA & NA & No & No \\
{\bf lockfree-3.1} & 93 & Symbolic Loop Condition & Memory Leak &  - & 2.17 & No & Yes \\
                        & & Dynamic Data Structure  & & & & & \\
{\bf lockfree-3.2} & 93 & Symbolic Loop Condition &  Memory Leak & 13.93 & 2.03 & Yes & Yes \\
                        & & Dynamic Data Structure  & & & & & \\
{\bf test-0019-2} & 30 & Dynamic Data Structure & Memory Leak & 0.07 & 2.02 & Yes & Yes \\
{\bf test-0102-2} & 95 & Symbolic Loop Condition & Memory Leak & 210.38 & 2.02 & Yes & Yes \\
                        & & Dynamic Data Structure  & & & & & \\
{\bf test-0137} &122 & Symbolic Loop Condition & OOB & 2.17 & 1.02 & Yes & Yes \\
                        & & Dynamic Data Structure  & & & & & \\
{\bf test-0158-2} & 19 & Aliasing & Memory Leak &  0.10 & 2.04 & Yes & Yes \\
                        & & Dynamic Data Structure  & & & & & \\
{\bf test-0158-3} & 23 & Aliasing & Double Free & 0.11 & 1.02 & Yes & Yes \\
                        & & Dynamic Data Structure  & & & & & \\
{\bf test-0220} & 65 & Symbolic Loop Condition & Memory Leak & 0.20 & 2.02 & Yes & Yes\\
                        & & Dynamic Data Structure  & & & & & \\
{\bf test-0232-1} &36 & Symbolic Loop Condition & Memory Leak & 0.78 & 2.02 & Yes & Yes\\
                       & & Dynamic Data Structure  & & & & & \\
{\bf test-0232-3} & 39 & Symbolic Loop Condition & Double Free & 1.55 & 1.02 & Yes & Yes\\
                       & & Dynamic Data Structure  & & & & & \\
{\bf test-0234-2} & 139 &  Symbolic Loop Condition & Memory Leak & 1.10 & 2.04 & Yes & Yes \\
                       & & Dynamic Data Structure  & & & & & \\
{\bf test-0235-1} & 153 &  Symbolic Loop Condition & Memory Leak & - & 2.02 & Yes & Yes \\
                       & & Dynamic Data Structure  & & & & & \\ 
{\bf test-0235-3} & 155& Symbolic Loop Condition & Use After Free &-  & 1.02 & No & Yes \\ 
                       & & Dynamic Data Structure  & & & & & \\
\bottomrule
\end{tabular} 
\end{footnotesize}
\end{table*}

\subsection{\bf RQ2: Can DESTINA Guide Forward Symbolic Execution?}
\label{sec:rq2}

DESTINA produces all valid program paths as long as the the fork limit and the in-cycle edge limit is not reached. However, this means that it may spend some of its time on resolving expressions for infeasible paths. This motivated us to combine MPBSE with 
DSE so that even if MPBSE does not reach the target it can collect some constraints 
during its exploration and guide DSE to help reach the target faster than the case without any guidance. So, to evaluate DESTINA's guidance capabilities we used Svcomp openssl control-flow benchmarks that included paths that can reach error locations. 
So, we set the target to the error locations and used a fork limit and in-cycle edge limit of 1. In this experiment, we compared DESTINA against KLEE 1.4.0 on which we built our tool.
Table \ref{table:buggygfse} shows that in 7 out of 11 cases DESTINA reduced the number of paths as well as the total time to reach the target. 
DESTINA achieves on average 4.36X reduction on the number of paths and 2.86X speedup. 
{\bf The results show that DESTINA can speedup Dynamic Symbolic Execution by inferring solutions that are likely to reach a given target.}

\begin{table*}[th!]
\caption{TSE and DSE on buggy {\tt SvComp openssl control-flow} benchmarks with reachable assertion failures. $X^{\textit{OOM}}$ denotes not reaching the target and terminating due to out of memory error after generating $X$ paths. Timeout (TO) is set to 600 secs.}
\label{table:buggygfse}
\centering
\begin{footnotesize}
\begin{tabular}{c|r|r|r|r|r|r|r|r|r|r|r|r} \toprule
{\bf Benchmark} & {\bf SLOC} & \multicolumn{2}{c|}{\bf Reached?} & \multicolumn{2}{c|}{\bf \# Paths to Target} & \multicolumn{4}{c}{\bf Time (secs)} & {\bf $|Replay|$}\\ \cline{3-11}
 & & {\bf DSE} & {\bf TSE} & {\bf DSE} & {\bf TSE} & {\bf DSE} & {\bf TSE Total} & {\bf MPBSE} & {\bf GFSE} & \\ \hline
{\bf s3\_clnt\_1.cil-2} & 557 & Yes & Yes & 206 & 48 & 23.67  & 5.00 & 1.03 & 4.97 & 4\\
{\bf s3\_clnt\_3.cil-2}	& 591 & Yes & Yes & 406	& 98 & 57.50 & 12.50 & 1.41 & 11.09 & 4\\
{\bf s3\_clnt\_4.cil-2}	& 562 & Yes & No & 209	& 222 & 8.50 & 12.00 & 0.17 & 11.83 & 0 \\
{\bf s3\_srvr\_1.cil-1}	& 628 & Yes & Yes & 81	& 16 & 6.00 & 2.50 & 1.34 & 1.16 & 5\\
{\bf s3\_srvr\_10.cil}	& 644 & Yes & No & 13	& - & 0.60 & TO & TO & TO & 5\\
{\bf s3\_srvr\_11.cil}	& 650 & No & Yes & $1270^{\textit{OOM}}$	& 189 & 220.00 & 40.67 & 1.37 & 39.30& 5\\
{\bf s3\_srvr\_12.cil}	& 716 & No & No & $844^{\textit{OOM}}$	& $1310^{\textit{OOM}}$	 & 128.00 & 300.03 & 1.10 & 298.83 & 5 \\
{\bf s3\_srvr\_13.cil}	& 658 & Yes & Yes & 306 & 99 & 57.00 & 19.67 & 1.12 & 18.55 & 5\\
{\bf s3\_srvr\_14.cil}	& 656 & Yes & Yes & 43 &	5 & 2.50 & 1.00 & 0.78 & 0.22 & 5\\
{\bf s3\_srvr\_2.cil-2}	& 624 & Yes & Yes & 81 &	15 & 5.50 & 2.00 & 1.13 & 0.70 & 5\\
{\bf s3\_srvr\_6.cil-1}	& 687 & Yes & Yes & 1	& 1 & 0.29 & 1.07 & 0.97 & 0.03 & 5\\ \bottomrule
\end{tabular} 
\end{footnotesize}
\end{table*}

%% file: relatedwork.tex
\section{Related Work}
\label{sec:relwork}

The weakest precondition computation  is a principled approach to backward reasoning. 
Automatic computation of weakest preconditions has been adapted to intra-procedural \cite{FLL02} and inter-procedural analysis \cite{CFS09} of Java programs
and to analysis of PHP code in \cite{ODL15}.  
The preconditions generated by these approaches can potentially include constraints belonging to undecidable theories such as nonlinear constraints for integers.
This issue has been dealt by using user assertions along with a theorem prover in \cite{FLL02}, by under-approximating the analysis in \cite{CFS09}, and by
generating over-approximate constraints in \cite{ODL15}. 
In \cite{GKW15}, weakest-preconditions computed over forward executed program segments of concurrent programs are used to summarize explored paths and to avoid redundant computations.
The backward reasoning mode in our approach does not aim to compute weakest preconditions. 
Instead, its goal is to define the expressions using the abstract address space while 
generating a backward symbolic execution tree.

Backward Bounded Model Checking and Dynamic Symbolic Execution are combined in BB-DSE \cite{BDM17} to answer infeasibility queries and to support deobfuscation of binaries. 
Our approach differs from this work in several dimensions: 1) BB-DSE answer infeasibility queries whereas our approach checks for feasibility while supporting precise detection of memory vulnerabilities. 2) BB-DSE performs bounded analysis and it may have both false positives and false negatives whereas our approach does not have any false positives. 


Three targeted symbolic execution approaches are presented in \cite{MK11} to solve the line reachability problem :  shortest-distance symbolic execution (SDSE), call-chain-backward symbolic execution (CCBSE), and Mix-CCBSE. SDSE  steers the execution towards the target by prioritizing branches that have the shortest distance to the target. CCBSE checks reachability of the target 
by turning it into  a multi-step reachability query, where the first query checks the reachability of the target from the entry point of the function that directly calls it, the second query checks the reachability of 
the callsite of the function directly calling the target from the entry point of its caller, and so on. Mix-CCBSE mixes SDSE and CCBSE to leverage the advantage of both approaches. 
CCBSE stitches the results of multiple under-constrained symbolic execution and, therefore, directly uses the memory modeling of forward symbolic execution. However, TSE employs a truly backward symbolic execution that introduces abstract memory objects. Through unification of the abstract objects with the concrete ones and early evaluation, TSE can identify infeasible paths even before reaching the entry point of some function.

Preconditioned symbolic execution \cite{ACH11} uses standard forward symbolic execution while restricting the exploration to paths that satisfy a provided precondition. 
Preconditioned symbolic execution relies on another analysis to derive the precondition 
whereas our approach automatically generates target relevant constraints.

The approach of \cite{DA14} is closest to our approach in that it performs backward analysis from the target and then validates the feasibility of the path to the target through forward execution. However, the approach analyzes Java programs, only handles constraints over integers, and mentions the use of a custom solver for object related constraints without providing the details. 

Postconditioned symbolic execution \cite{YYG15,YYG18} keeps track of the postconditions at the symbolic branch locations to avoid the reexploration of
covered path suffixes. 
Our approach, on the other hand, avoids the reexploration 
of path prefixes in two ways. The first one eliminates the exploration of infeasible path prefixes due to early evaluation of the infeasible path constraints.
The second one eliminates the exploration of feasible path prefixes by terminating the exploration upon detecting the first feasible path when answering a reachability query.

BSELF \cite{CS21} extends Backward Symbolic Execution (BSE) with dynamically generated inductive loop invariants  to  verify integer programs. It cannot handle loop bodies that read inputs. 
Although TSE cannot verify correctness of code as is the case in standard dynamic symbolic execution, 
TSE has advantage over BSELF for buggy code or reachable targets as it is not limited to integer programs.

Recently, machine learning models of vulnerable code have been used in \cite{ZCW22} to direct forward symbolic execution. This is orthogonal to our approach and machine learning models can also be integrated to our approach to direct the backward execution mode.


%% file: conclusions.tex
\section{Conclusions}
\label{sec:conc}

We presented a novel targeted symbolic execution approach that uses a mix of forward and backward passes and precisely computes points-to information at the byte granularity. 
Our tool, DESTINA, has been implemented using KLEE. DESTINA can both detect memory errors and collect useful information to steer standard symbolic execution towards a target location. Our results show that 
DESTINA can detect five types of memory errors and achieves on average 4X reduction in the number of paths explored before reaching the target. In future work, we will extend DESTINA with bug specific heuristics.